\documentstyle[12pt,times]{article}
\begin{document}
\def \beq{\begin{equation}}
\def \eeq{\end{equation}}
\def \bea{\begin{eqnarray}}
\def \eea{\end{eqnarray}}
\def \ee {e^+e^-}
\def \ra {\rightarrow}
\def \g {\gamma}
\def \cvg {c_v^{\gamma}}
\def \cag {c_a^{\gamma}}
\def \cvz {c_v^Z}
\def \caz {c_a^Z}
\def \cdg {c_d^{\gamma}}
\def \cdz {c_d^Z}
\def \cdgz {c_d^{\gamma,Z}}
\def \t{$t\;$}
\def \toverline{$\overline t$}
\def \tbar{$\overline t\;$}
\def \el{E_l}
\def \ttt{\theta_t}
\def \tht{\theta_t}
\def \thl{\theta_l}
\def \thetal{\theta_l}
\def \phil{\phi_l}
\def \CP{$CP$}
\def \f{\frac}
\def \o{\overline}
\def \n{\noindent}
\def \half{\frac{1}{2}}
\def \Re{{\rm Re}}
\def \re{{\rm Re}}
\def \Im{{\rm Im}}
\def \im{{\rm Im}}
\def \bb{\bibitem}
\begin{flushright}
hep-ph/0105318
\end{flushright}
\begin{center}
{\large \bf \boldmath Decay-lepton angular asymmetries in polarized
$e^+e^- \ra t\o{t}$ as a measure of $CP$-violating dipole couplings
of the top quark } 
\vskip .5cm 
{\large Saurabh D. Rindani\footnote{Email address: \tt saurabh@prl.ernet.in}} 
\vskip .25cm 
{\it Theory Group, Physical Research Laboratory \\
Navrangpura, Ahmedabad 380009, India} 
\vskip 1cm 
{\small \bf Abstract}
\end{center}
\vskip .2cm
\def \tt {t\overline {t}}
\begin{quote}
{\small
In the presence of an electric dipole coupling of $\tt$
to a photon, and an analogous ``weak" dipole coupling to the $Z$,
$CP$ violation in the process $\ee \ra \tt$ leads to the polarization
of the top and anti-top. This polarization can be analyzed by
studying the angular distributions of decay charged
leptons when the top or anti-top decays leptonically. We
have obtained analytic expressions for these distributions when
either $t$ or $\overline t$ decays leptonically, including ${\cal O}
(\alpha_s)$ QCD corrections in the soft-gluon approximation. The angular
distributions are insensitive to anomalous interactions in top decay. 
We study two types
of simple $CP$-violating polar-angle asymmetries and two azimuthal asymmetries
which do not need the
full reconstruction of the $t$ or $\o t$. We
have evaluated independent 90\% CL limits that may be obtained on the real and
imaginary parts of the electric and weak dipole couplings at a linear collider
operating at $\sqrt{s}=500$ GeV with integrated luminosity 200 fb$^{-1}$ and 
also at $\sqrt{s}=1000$ GeV with integrated luminosity 1000 fb$^{-1}$. 
The effect of longitudinal beam polarization has been included. 
}
\end{quote}
\vskip .5cm
\section{Introduction}
An $e^+e^-$  linear collider operating at centre-of-mass (cm)
energy of 500 GeV or higher and with an integrated luminosity of several
hundred inverse femtobarns should be able to study with precision
various properties of the top quark.

While the standard model (SM) predicts $CP$ violation outside
the \mbox{$K$-,}
$D$- and $B$-meson systems to be unobservably small,
in some extensions of SM, $CP$ violation might be considerably
enhanced, especially in the presence of a heavy top quark.  In
particular,
$CP$-violating electric dipole form factor of the top quark, and
the analogous $CP$-violating ``weak" dipole form factor in the $\tt$
coupling
to $Z$, could be enhanced.  These $CP$-violating form factors could
be
determined in a model-independent way at high energy $\ee$ linear
colliders,
where $\ee \ra \tt$ would proceed through virtual $\g$ and $Z$
exchange.

Since a heavy top quark with a mass of the order of 175 GeV
decays before it hadronizes \cite{heavytop}, it has
been suggested \cite{toppol} that top polarization asymmetry in
$\ee \ra \tt$ can be used to determine the $CP$-violating dipole
form factors, since polarization information would be retained in
the decay product distribution.  There have been several proposals
in which the $CP$-violating dipole couplings could be measured in
decay momentum correlations or asymmetries  with or without beam
polarization. For a review see \cite{sonirev}.

In this context it is important to note that top polarization can
only be studied using top decay. However, for the information
from decay distributions to reflect correctly top polarization,
the decay amplitudes for various top polarization states have to
be known accurately. In particular, if there are any anomalous
effects in the decay process, they have to be known accurately.
Alternatively, the decay distributions chosen for the study have
to be insensitive to anomalous effects in the decay process. The
single-lepton angular distributions that we discuss in this work
satisfy the latter condition -- they accurately reflect the
polarization of the top quark resulting from the production
process, while one can continue to use SM in the decay process.

In this paper we update some suggestions made
\cite{pou1,pou2,pou3} for the measurement of top dipole moments in
$\ee \ra \tt$ using angular asymmetries of the charged lepton
produced in the semi-leptonic decay of one of $t$ and \tbar, while
the other decays hadronically. The improvements included in the
update are several. Firstly, there is now a better idea of
luminosities possible at a future linear collider like the proposed JLC. 
Together with
updated values of beam polarization now considered feasible, the
estimates of possible limits on dipole moments would be more
realistic. Secondly, ${\cal O}(\alpha_s)$ QCD corrections in the
soft-gluon approximation have been now been included. Thirdly, an
assumption made in earlier work \cite{pou1,pou2,pou3}, that CP
violation in top decay could be neglected, has been re-examined in
light of recent work \cite{hioki,sdr1}. It turns out that for
angular asymmetries of the charged lepton considered here, CP
violation in the decay (or for that matter even arbitrary
\CP-conserving modifications of the $tbW$ vertex) has no effect,
if the $b$-quark mass is neglected. Thus the estimates in earlier
work have been improved and put on sounder theoretical footing.

Earlier proposals have considered a variety of CP-violating
observables, with varying sensitivies. These include, in addition to angular
asymmetries, also vector and tensor correlations \cite{bern,cuy}, 
and expectation values of 
optimal variables \cite{atwood}. (For a discussion on
relative sensitivities of some variables, see \cite{lietti}). We
have chosen certain angular asymmetries here which have some
advantages over others, even though they may not be the most
sensitive ones. The advantages are: (i) Our asymmetries are in the
laboratory frame, making them directly observable. (ii) They
depend on final state momenta, rather than on top polarization.
Polarization is measured only indirectly through the decay
distributions. We therefore concentrate only on actual
decay-lepton distributions, which are the simplest to observe.
(iii) The observables we choose either do not depend on precise
determination of energy and momentum of top quarks, or, in case of
azimuthal asymmetries of the lepton, depend minimally on the top
momentum direction for the sake of defining the coordinate axes.
This has the advantage of higher accuracy. (iv) As stated before,
leptonic angular distribution is free from background from CP
violation in top decay, and gives a direct handle on anomalous
couplings in top production. (v) The polar-angle asymmetries we
consider can be obtained in analytical form, which is useful for
making quick computations. It is possible to get analytical forms
for certain azimuthal asymmetries as well, provided no angular
cuts are imposed. (vi) The asymmetries considered here are rather
simple conceptually, and hopefully, also from the practical
measurement point of view.

Our single-lepton asymmetries have another obvious advantage, that
since either $t$ or $\o{t}$ is allowed to decay hadronically,
there is a gain in statistics, as compared to double-lepton
asymmetries.

Our results are based on fully analytical calculation of single
lepton distributions in the production and subsequent decay of
$\tt$. We present fully differential angular distribution as well as the
distribution in the polar angle of the lepton with respect to the
beam direction in the centre-of-mass (cm) frame for arbitrary
longitudinal beam polarizations. These distributions for the
standard model (SM) were first obtained by Arens and Sehgal
\cite{arens}. Distributions including the effect of CP violation
only in production were obtained in \cite{pou2,pou3}, whereas,
with all anomalous effects included in the $\gamma t\overline{t}$
and $Zt\overline{t}$ vertices, as well as decay $tbW$ vertex were
obtained in \cite{hioki,sdr1}. Angular distributions in SM with
${\cal O}(\alpha_s)$ QCD corrections in the soft-gluon
approximation were obtained in \cite{sga}. The distributions
including anomalous effects in both top production and decay, and
including ${\cal O}(\alpha_s)$ QCD corrections in the soft-gluon
approximation are presented here for the first time. While QCD corrections to
$e^+e^- \ra t\o{t}$ are substantial, to the extent of about 30\% at $\sqrt{s} =
500$ GeV, their effect on leptonic angular distributions is much smaller
\cite{sga}. The main effect on the results will be to the sensitivity, through
the $1/\sqrt{N}$ factor, where $N$ is the number of events.

The rest of the paper is organized as follows. In Sec. 2, we
describe the cal\-cula\-tion of the decay-lepton angular
dist\-ribu\-tion from a decaying  $t$ or $\o t$ in $\ee \ra \tt$.
In Sec. 3 we describe $CP$-violating asymmetries. Numerical
results are presented in Sec. 4, and Sec. 5 contains our
conclusions. The Appendix contains certain expressions which are
too lengthy to be put in the main text.

\section{Calculation of  lepton  angular distributions}

We describe in this section the calculation of $l^+\; (l^-)$
distribution in
$\ee \ra \tt$ and the subsequent decay $t \ra b l^+ \nu_l\;
(\overline{t} \ra
\overline{b} l^- \overline{\nu_l})$.  We adopt the narrow-width
approximation for
$t$ and $\overline{t}$, as well as for $W^{\pm}$ produced in
$t,\;\overline{t}$
decay.

  We assume the top quark couplings to $\g$ and $Z$ to be
given by the vertex factor  $ie\Gamma_\mu^j$, where
\beq
\Gamma_\mu^j\;=\;c_v^j\,\g_\mu\;+\;c_a^j\,\g_\mu\,\g_5\;+
\;\f{c_d^j}{2\,m_t}\,i\g_5\,
(p_t\,-\,p_{\overline{t}})_{\mu},\;\;j\;=\;\g,Z,
\eeq
with
\bea
\cvg&=&\f{2}{3},\:\;\;\cag\;=\;0, \nonumber \\
\cvz&=&\f {\left(\f{1}{4}-\f{2}{3} \,x_w\right)}
{\sqrt{x_w\,(1-x_w)}},
 \\
\caz&=&-\f{1}{4\sqrt{x_w\,(1-x_w)}}, \nonumber \eea and
$x_w=sin^2\theta_w$, $\theta_w$ being the weak mixing angle. In
addition to the SM couplings $c^{{\g},Z}_{v,a}$ we have introduced
the $CP$-violating electric and weak dipole form factors,
$e\cdg/m_t$ and $e\cdz/m_t$, which are assumed small.  Use has
also been made of the Dirac equation in rewriting the usual dipole
coupling $\sigma_{\mu\nu}(p_t+p_{\overline{t}})^{\nu}\g_5$ as
$i\g_5(p_t-p_{\overline{t}})_{\mu}$, dropping small corrections to
the vector and axial-vector couplings. We will work in the
approximation in which we keep only linear terms in $\cdg$ and
$\cdz$. Addition of other \CP-conserving form factors will not
change our results in the linear approximation.

To include ${\cal O}(\alpha_s)$ corrections in the soft-gluon approximation (SGA), we need to modify the above vertices, as explained in \cite{sga}.
These modified vertices are given by
\beq\label{gvert}
\Gamma_{\mu}^{\gamma} =  c_v^{\gamma} \gamma_{\mu} + \left[ c_M^{\gamma}
 +
i\g_5\, c_d^{\gamma} \right]\,
\frac{(p_t - p_{\overline{t}})_{\mu}}{2 m_t} ,
\eeq
\beq\label{Zvert}
\Gamma_{\mu}^{Z} =  c_v^{Z} \gamma_{\mu} +  c_a^{Z}
\gamma_{\mu}\gamma_5
  + \left[ c_M^{Z}+
i\g_5 \, c_d^Z\right]\, \frac{(p_t - p_{\overline{t}})_{\mu}}{2
m_t} , \eeq where \beq c_v^{\gamma} = \f{2}{3} (1+A), \eeq \beq
c_v^Z = \frac{1}{\sin\theta_W \cos\theta_W} \left( \frac{1}{4} -
\frac{2}{3} \sin^2\theta_W\right) (1+A), \eeq \beq c_a^{\gamma}=0,
\eeq \beq c_a^Z = \frac{1}{\sin\theta_W \cos\theta_W}
\left(-\frac{1}{4}\right) (1+A+2 B), \eeq \beq c_M^{\gamma} =
\frac{2}{3} B, \eeq \beq c_M^Z = \frac{1}{\sin\theta_W
\cos\theta_W} \left( \frac{1}{4} - \frac{2}{3}
\sin^2\theta_W\right) B. \eeq The form factors $A$ and $B$ are
given to order $\alpha_s$ in SGA (see, for example, \cite{kodaira,tung} by \bea\label{A} {\rm
Re} A &=& \hat{\alpha}_s \left[ \left( \frac{1+\beta^2}{\beta}
\log \frac{1+\beta}{1-\beta} - 2\right)\log\frac{4 \omega^2_{\rm
max}}{m_t^2} - 4 \right.\nonumber \\ && \left. +
\frac{2+3\beta^2}{\beta}\log\frac{1+\beta}{1-\beta} +
\frac{1+\beta^2}{\beta} \left\{ \log\frac{1-\beta}{1+\beta}\left(
3 \log\frac{2\beta}{1+\beta}\right.\right.\right. \nonumber \\ &&
\left.\left.\left. + \log\frac{2\beta}{1-\beta} \right) + 4 {\rm
Li}_2 \left(\frac{1-\beta}{1+\beta}\right) +
\frac{1}{3}\pi^2\right\}\right], \eea \beq\label{B} {\rm Re}
B=\hat{\alpha}_s \frac{1-\beta^2}{\beta} \log
\frac{1+\beta}{1-\beta}, \eeq \beq {\rm Im} B = - \hat{\alpha}_s
\pi \frac{1-\beta^2}{\beta}, \eeq where
$\hat{\alpha}_s=\alpha_s/(3\pi)$, $\beta = \sqrt{1-4 m_t^2/s}$,
and Li$_2$ is the Spence function. ${\rm Re} A$ in eq. (\ref{A})
contains the effective form factor for a cut-off $\omega_{\rm
max}$ on the gluon energy after the infrared singularities have
been cancelled between the virtual- and soft-gluon contributions
in the on-shell renormalization scheme. Only the real part of the
form factor $A$ has been given, because the contribution of the
imaginary part is proportional to the $Z$ width, and hence
negligibly small \cite{kodaira,ravi}. The imaginary part of $B$,
however, contributes to azimuthal distributions.

The helicity amplitudes for $\ee \ra \g^*,Z^* \ra \tt$ in the cm
frame, including $\cdgz$ and $c_M^{\g ,Z}$ couplings, have been
given in \cite{asymm} (see also Kane {\it et al.}, ref.
\cite{toppol}).

We write the contribution of a general $tbW$ vertex to $t$ and \tbar decays
as
\bea\label{tbw}
\Gamma^{\mu}_{tbW}& =& -\f{g}{\sqrt{2}}V_{tb} \o{u}(p_b)
\left[\gamma^{\mu}(f_{1L}
P_L+f_{1R}P_R)\f{}{} \right. \nonumber \\
&& \left. - \f{i}{m_W} \sigma^{\mu\nu} (p_t - p_b)_{\nu}
 (f_{2L}P_L+f_{2R}P_R) \right] u(p_t),
\eea
\bea\label{tbarbw}
\o{\Gamma}^{\mu}_{tbW}& =& -\f{g}{\sqrt{2}}V_{tb}^* \o{v}(p_{\o{t}})
\left[\gamma^{\mu}(\o{f}_{1L} P_L + \o{f}_{1R} P_R )\f{}{} \right. \nonumber \\
&& \left. -
\f{i}{m_W} \sigma^{\mu\nu} (p_{\o{t}} - p_{\o{b}})_{\nu} (\o{f}_{2L}P_L+\o{f}
_{2R}P_R) \right]
v(p_{\o{b}}),
\eea
where $P_{L,R} =\half (1\pm \gamma_5)$, and $V_{tb}$ is the
Cabibbo-Kobayashi-Maskawa matrix element, which we take to be equal to one.
If CP is conserved, the form factors $f$ above obey the relations
\beq
f_{1L}=\o{f}_{1L};\;\; f_{1R}=\o{f}_{1R},
\eeq
and
\beq
f_{2L}=\o{f}_{2R};\;\; f_{2R}=\o{f}_{2L}.
\eeq
Like $c_d^{\gamma}$ and $c_d^Z$ above, we will also treat $f_{2L,R}$ and
$\o{f}_{2L,R}$ as small, and retain only terms linear in them.
For the form factors $f_{1L}$ and $\o{f}_{1L}$, we retain their SM values,
viz., $f_{1L} = \o{f}_{1L} = 1$. $f_{1R}$ and $\o{f}_{1R}$ do not contibute in
the limit of vanishing $b$ mass, which is used here. Also, $f_{2L}$ and
$\o{f}_{2R}$ drop out in this limit.

The
helicity amplitudes for
\[t \ra b W^+,\;
\;W^+ \ra l^+ \nu_l\]
and\[\overline{t} \ra \overline{b}W^-,\;\; W^- \ra
l^-\overline{\nu_l}\]
in the respective rest frames of $t$, $\overline{t}$,
in the limit that all masses
except the top mass are neglected,
are given in ref. \cite{sdr1}.

Combining the production and decay amplitudes in the narrow-width
approximation for $t,\overline{t},W^+,W^-$, and using appropriate
Lorentz
boosts to calculate everything in the $\ee$ cm frame, we get the
$l^+$
and $l^-$ angular distributions for the case of $e^-$, $e^+$ with
polarization
$P_e$, $P_{\o e}$ to be:
\bea\label{triple}
\lefteqn{\f{d^3\sigma^{\pm}}{d\cos\ttt d\cos\thl d\phil} = \f{3\alpha^2\beta
m_t^2}{8s^{2}}B_tB_{\o{t}} \f{1}{(1-\beta\cos\theta_{tl})^3}}
\nonumber\\
&\times &\left[{\cal  A} (1-\beta\cos\theta_{tl})
+{\cal B}^{\pm} (cos\theta_{tl} - \beta )\right. \nonumber \\
&&\left.
+{\cal  C}^{\pm} (1-\beta^2)\sin\tht\sin\thl (\cos\tht\cos\phil -
\sin\tht\cot\thl )\right. \nonumber \\
&&\left. +{\cal D}^{\pm} (1-\beta^2) \sin\tht\sin\thl\sin\phil \right],
\eea
where $\sigma^+$ and $\sigma^-$ refer respectively to $l^+$ and
$l^-$ distributions, with the same notation for the kinematic
variables of
particles and antiparticles.  Thus, $\theta_t$, is the polar angle of
$t$  (or \tbar), and $\el,\;\thetal,\;\phil$ are the energy, polar
angle and azimuthal angle of $l^+$ (or $l^-$).
All the angles are now in the cm frame, with the $z$ axis chosen
along the $e^-$ momentum, and the $x$ axis chosen in the plane
containing the $e^-$ and $t$ directions.
$\theta_{tl}$ is the angle between
the $t$ and $l^+$ directions (or $\overline t$ and $l^-$ directions).
$\beta$
is the $t$ (or $\o t$) velocity: \(\beta=\sqrt{1-4m_t^2/s}\), and
$\gamma = 1/\sqrt{1-\beta^2}$. $B_t$ and $B_{\o t}$ are respectively the
branching ratios of $t$ and
$\o t$ into the final states being considered.

The coefficients ${\cal A}^{\pm}$, ${\cal B}^{\pm}$, ${\cal C}^{\pm}$ and ${\cal D}^{\pm}$ are
given  by
\bea
{\cal A} & = & A_0 + A_1 \cos\tht + A_2 \cos^2\tht , \\
{\cal B}^{\pm} & = & B_0^{\pm} + B_1 \cos\tht + B_2^{\pm} \cos^2\tht , \\
{\cal C}^{\pm} & = & C_0^{\pm} + C_1^{\pm} \cos\tht , \\
{\cal D}^{\pm} & = & D_0^{\pm} + D_1^{\pm} \cos\tht ,
\eea
The quantities $A_i$, $B_i^{\pm}$, $C_i^{\pm}$ and $D_i^{\pm}$ occurring
in the above
equation are functions of the masses, $s$, the degrees of $e$ and $\o
e$  polarization ($P_e$ and $P_{\o e}$), and the coupling constants.
They are listed in the Appendix.

It should be emphasized that, as shown in \cite{hioki,sdr1}, the distribution
in (\ref{triple}) does not depend on anomalous effects in the $tbW$ vertices 
(\ref{tbw}) and (\ref{tbarbw}). Thus even ${\cal O}(\alpha_s)$ QCD corrections
to the $tbW$ vertices would not be felt in (\ref{triple}).

To obtain the single-differential polar-angle distribution, we integrate over 
$\phi$ from 0 to $2 \pi$, and finally over cos$\,\theta_t$
from $-1$ to +1.  The final result is

\bea\label{cldist}
\lefteqn{
\f{d\sigma^{\pm}}{d\cos\theta_l}=\frac{3\pi\alpha^2}{32s}
B_t B_{\overline t} \beta
\left\{4A_0
\mp 2A_1\left(\f{1-\beta^2}{\beta^2} \log\f{1+\beta}{1-\beta}-
\f{2}{\beta}\right) \cos\theta_l \right.}\nonumber \\
&&\left. + 2A_2 \left(
\f{1-\beta^2}{\beta^3}\log\f{1+\beta}{1-\beta}
(1-3\cos^2\theta_l) \right. \right. \nonumber \\
&& \left. \left. - \f{2}{\beta^2} (1-3\cos^2\theta_l-\beta^2+2 \beta^2
\cos^2\theta_l) \right) \right. \nonumber \\
&&\left. \pm 2B_1\f{1-\beta^2}{\beta^2}  \left(
\f{1}{\beta}\log\f{1+\beta}{1-\beta} -
2 \right) \cos\theta_l \right. \nonumber \\
&&  \left. + B_2^{\pm} \f{1-\beta^2}{\beta^3}
\left( \f{\beta^2-3}{\beta} \log\f{1+\beta}{1-\beta} + 6 \right)
(1-3\cos^2\theta_l) \right. \nonumber \\
&&\left. \pm
2C_0^{\pm}\f{1-\beta^2}{\beta^2} \left( \f{1-\beta^2}{\beta}
\log\f{1+\beta}{1-\beta} - 2 \right) \cos\theta_l
\right. \nonumber \\
&& \left. - C_1^{\pm}\f{1-\beta^2}{\beta^3} \left( \f{3(1-\beta^2)}{\beta}
\log\f{1+\beta}{1-\beta} -2(3-2\beta^2)\right) (1-3
\cos^2\theta_l) \right\}.
\eea
This is the same expression as in \cite{pou2} and \cite{sdr1}. However, the
significance of the functions $A_i$, $B_i$, $C_i$ and $D_i$ is different in
each case.

We now proceed to a discussion of \CP-odd asymmetries resulting from the use of
the above distributions.

\section{$CP$-violating angular asymmetries}

We will work with two different types of asymmetries, one which does not
depend on the azimuthal angles
angle of the decay lepton, so that the azimuthal angle is fully integrated over,
and the
other dependent on the azimuthal angle. In all cases, we assume a cut-off of
$\theta_0$ on the forward and backward directions of the charged lepton.
Some cut-off on the forward and backward angles is certainly needed from an
experimental point of view; we furthermore exploit the cut-off to optimize the
sensitivity.

In the first case, namely polar asymmetries,
we define two independent \CP-violating asymmetries, which depend on
different
linear combinations of Im$\cdg$ and Im$\cdz$ . (It is not
possible to define \CP-odd
 quantities which determine Re$\cdgz$ using single-lepton polar
distributions, as can
be seen from the expression for the \CP-odd combination
$\frac{d\sigma^+}{d\cos\theta_l}(\theta_l)
-\frac{d\sigma^-}{d\cos\theta_l}(\pi-\theta_l)$). One is simply the
total
lepton-charge asymmetry, with a cut-off of $\theta_0$ on the forward
and
backward directions:
\beq
A_{ch}(\theta_0)=\frac{
{\displaystyle      \int_{\theta_0}^{\pi-\theta_0}}d\theta_l
{\displaystyle          \left( \frac{d\sigma^+}{d\theta_l}
        -   \frac{d\sigma^-}{d\theta_l}\right)}}
{
{\displaystyle      \int_{\theta_0}^{\pi-\theta_0}}d\theta_l
{\displaystyle          \left( \frac{d\sigma^+}{d\theta_l} +
\frac{d\sigma^-}{d\theta_l}\right)}}.
\eeq
The other is the lept\-onic forward-backward asy\-mmetry com\-bined
with charge asy\-mmetry, again with the angles within $\theta_0$ of
the forward and back\-ward directions excluded:
\beq
A_{fb}(\theta_0)= \frac{ {\displaystyle
\int_{\theta_0}^{\frac{\pi}{2}}}d\theta_l {\displaystyle
\left( \frac{d\sigma^+}{d\theta_l} +
\frac{d\sigma^-}{d\theta_l}\right)} {\displaystyle
- \int^{\pi-\theta_0}_{\frac{\pi}{2}}}d\theta_l {\displaystyle
\left( \frac{d\sigma^+}{d\theta_l} +    \frac{d\sigma^-}{d\theta_l}
\right)}}
{
{\displaystyle      \int_{\theta_0}^{\pi-\theta_0}}d\theta_l
{\displaystyle          \left( \frac{d\sigma^+}{d\theta_l} +
\frac{d\sigma^-}{d\theta_l}\right)}}.
\eeq

Analytic expressions for both these aymmetries may be easily obtained using
(\ref{cldist}), and are not displayed here explicitly.

We note the fact that ${\cal A}_{ch}(\theta_0)$ vanishes for
$\theta_0=0$. This implies that the $CP$-violating charge asymmetry
does not exist unless a cut-off is imposed on the lepton production
angle. ${\cal A}_{fb}(\theta_0)$, however, is nonzero for
$\theta_0=0$.

We now define angular asymmetries of the second type, which depend on the
range of the
azimuthal angle $\phi_l$ of the charged lepton. These are called the up-down
and left-right asymmetries, and depend respectively on the real and imaginary
parts of the dipole couplings.

The up-down asymmetry [5] is defined by
\beq
A_{ud}(\theta_0)=\f{1}{2\,\sigma (\theta_0)}\int_{\theta_0}^{\pi-\theta_0}
\left[
\f{d\,\sigma^+_{\rm up} } {d\,\theta_l}
-\f{d\,\sigma^+_{\rm down} } {d\,\theta_l}
+
\f{d\,\sigma^-_{\rm up}} {d\,\theta_l}
-\f{d\,\sigma^-_{\rm down} } {d\,\theta_l}
\right] {d\,\theta_l} ,
\eeq
where
\beq\label{sigma0}
\sigma (\theta_0) = \int_{\theta_0}^{\pi-\theta_0} \f{d\,\sigma }
{d\,\theta_l}d\,\theta_l
\eeq
is the SM cross section for the semi-leptonic final state, with a forward and
backward cut-off of $\theta_0$ on $\thl$.
Here up/down refers to
$(p_{l^{\pm}})_y\;\raisebox{-1.0ex}{$\stackrel{\textstyle>}{<}$}\;0,\;
\:(p_{l^{\pm}})_y$ being the $y$
component of $\vec{p}_{l^{\pm}}$ with respect to a coordinate system
chosen in the $e^+\,e^-$ center-of-mass (cm) frame so that the
$z$-axis is along $\vec{p}_e$, and the $y$-axis is along
$\vec{p}_e\,\times\,\vec{p}_t$.  The $t\bar{t}$ production
plane is thus the $xz$ plane.  Thus, ``up" refers to the range $0<\phi_l<\pi$,
and ``down" refers to the range $\pi<\phi_l<2\pi$.

The left-right asymmetry is defined by
\beq
A_{lr}(\theta_0)=\f{1}{2\,\sigma (\theta_0)}\int_{\theta_0}^{\pi-\theta_0}
\left[
\f{d\,\sigma^+_{\rm left} } {d\,\theta_l}
-\f{d\,\sigma^+_{\rm right} } {d\,\theta_l}
+
\f{d\,\sigma^-_{\rm left}} {d\,\theta_l}
-\f{d\,\sigma^-_{\rm right} } {d\,\theta_l}
\right] {d\,\theta_l} ,
\eeq
Here left/right refers to
$(p_{l^{\pm}})_x\;\raisebox{-1.0ex}{$\stackrel{\textstyle>}{<}$}\;0,\;
\:(p_{l^{\pm}})_x$ being the $x$
component of $\vec{p}_{l^{\pm}}$ with respect to the coordinate system
system defined above.
Thus, ``left" refers to the range $-\pi /2<\phi_l<\pi /2$,
and ``right" refers to the range $\pi /2<\phi_l<3\pi /2$.

Analytic expressions for the up-down and left-right symmetry are not available
for nonzero cut-off in $\theta_l$. Hence, the angular integrations have been
done numerically in what follows.

Two other asymmetries were defined in \cite{pou1}, which helped to disentangle
the two dipole couplings from each other. However, we do not discuss these
here. Instead, we will assume that the electron beam polarization can be made
to change sign to give additional observable quantities to enable this
disentanglement.

All these asymmetries are a measure of $CP$ violation in the unpolarized
case and in the case when polarization is present, but
$P_e=-P_{\overline{e}}$.  When $P_e\neq -P_{\overline{e}}$, the
initial state is not invariant under $CP$, and therefore
$CP$-invariant interactions can contribute to the asymmetries.
However, to leading order in $\alpha$, these $CP$-invariant
contributions vanish in the limit $m_e=0$.  Order-$\alpha$ collinear
helicity-flip photon emission can give a $CP$-even contribution.
However, this background has been estimated in \cite{back}, and found to be
negligible for certain \CP-odd correlations for the kind of luminosities
under consideration. It has also been estimated for $A_{fb}$ and $A_{ch}$, and
again found negligible \cite{thesis}. The background is zero in the case of
$A_{ud}$ \cite{thesis}. It is expected that the background will also be 
negligible for $A_{lr}$ thought it has not been calculated explicitly. 

\section{Numerical Results}

In this section we describe the numerical results for the calculation
of 90\% confidence level (CL) limits that could be put on Re$\cdgz$ and 
Im$\cdgz$ using the asymmetries described in the previous sections.

We look at only semileptonic final states. That is to say, when $t$
decays leptonically, we assume $\o t$ decays hadronically, and {\it
vice versa}. We sum over the electron and muon decay channels. Thus,
$B_tB_{\o t}$ is taken to be $2/3\times2/9$.

We have considered unpolarized beams, as well as the case when the electron
beam has a longitudinal polarization of 90\%, 
either left-handed or right-handed. We
have also considered the possibility of two runs for identical time-spans
with the polarization reversed in the second run. The positron beam is assumed
to be unpolarized.

We assume an integrated luminosity of 200 fb$^{-1}$ for a cm energy of 500 GeV,
and an integrated luminosity of 1000 fb$^{-1}$ for a cm energy of 1000 GeV. The
limits for higher luminosities can easily be obtained by scaling down the
limits presented here by the square root of the factor by which the luminosity
is scaled up.

We use the parameters $\alpha = 1/128$, $\alpha_s(m_Z^2)=0.118$, $m_Z= 91.187$
GeV, $m_W=80.41$ GeV, $m_t=175$ GeV and $\sin^2\theta_W=0.2315$. We have used,
following \cite{kodaira}, a gluon energy cut-off of 
$\omega_{\rm max}=(\sqrt{s}-2m_t)/5$. While qualitative results would be
insensitive, exact
quantitative results would of course depend on the choice of cut-off.

\begin{figure}[ptb]
\begin{center}
\input{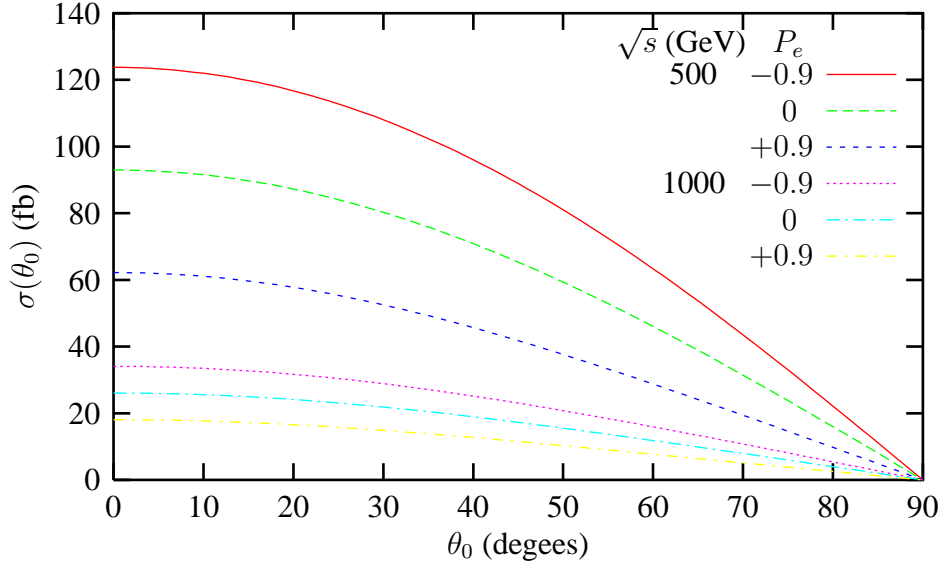}
\caption{\small
The  SM cross
section for decay leptons in the process $e^+e^- \rightarrow t\overline{t}$
plotted as a function of the cut-off $\theta_0$ on the lepton polar angle
in the forward and backward directions for $e^-$ beam longitudinal
polarizations $P_e=-0.9,0,+0.9$ and for values of total cm energy
$\sqrt{s}= 500$
GeV and $\sqrt{s}=1000$ GeV.}
\label{graph1}
\end{center}
\end{figure}

Fig. \ref{graph1}  shows the SM cross section $\sigma (\theta_0)$, defined in
eq. (\ref{sigma0}), for 
$t$ or $\o{t}$ production, followed by its semileptonic decay, with a cut-off
$\theta_0$ on the lepton polar angle, plotted against $\theta_0$ for the two 
choices of $\sqrt{s}$ and for different electron beam polarizations.

Tables 1-5
 show the results on the limits obtainable for each of these
possibilities. In all cases, the value of the cut-off $\theta_0$ has been
chosen to get the best sensitivity for that specific item.

\begin{table}
\begin{center}
\begin{tabular}{|c|c|c|c|c|c|c|c|}
\hline
&&
\multicolumn{3}{|c|}{$A_{ch}$}&\multicolumn{3}{|c|}{$A_{fb}$}\\
\cline{3-8}
$\sqrt{s}$ (GeV) &$P_e$ & $\theta_0$ & Im$c_d^{\gamma}$ &
 Im$c_d^Z$ &$\theta_0$ & Im$c_d^{\gamma}$ & Im$c_d^Z$ \\
\hline
    &  0    &64$^{\circ}$ & $ 0.084$ & $ 0.49$ & $10^{\circ}$ & 0.086 & 0.95\\
500 &$+0.9$ &64$^{\circ}$ & $ 0.081$ & 0.17   & $10^{\circ}$ & 0.075 &$ 0.15$\\
    &$-0.9$ &64$^{\circ}$ & $ 0.083$ & $ 0.14$ & $10^{\circ}$ & 0.093 & 0.15 \\
\hline
    & 0      &64$^{\circ}$ & $ 0.029$ & $ 0.18$ & $10^{\circ}$ &0.032&0.36 \\
1000& $+0.9$ &64$^{\circ}$ & $ 0.028$ & 0.061 &   $10^{\circ}$ &0.028&$ 0.058$\\
    & $-0.9$ &64$^{\circ}$ & $ 0.028$ &$ 0.047$&  $10^{\circ}$ &0.034&$ 0.058$\\
\hline
\end{tabular}
\caption{Individual 90\% CL limits on dipole couplings obtainable from
$A_{ch}$ and
$A_{fb}$ for $\sqrt{s}=500$ GeV with integrated luminosity 200 fb$^{-1}$
and for $\sqrt{s}=1000$ GeV with integrated luminosity 1000 fb$^{-1}$ for
different electron beam polarizations $P_e$. Cut-off $\theta_0$ is chosen to
optimize the sensitivity.}
\end{center}
\end{table}

In Table 1 we give the 90\% confidence level (CL) limits that can be 
obtained on Im$c_d^{\g}$ and
Im$c_d^Z$, assuming one of them to be nonzero, the other taken to be vanishing.
The limit is defined as the value of Im$\cdg$ or Im$\cdz$ for which the
corresponding asymmetry $A_{ch}$ or $A_{fb}$ becomes equal to $1.64/\sqrt{N}$,
where $N$ is the total number of events.

\begin{table}
\begin{center}
\begin{tabular}{|c|c|c|c|}
\hline
$\sqrt{s}$ (GeV)  & $\theta_0$ & Im$c_d^{\gamma}$ & Im$c_d^Z$\\
\hline
500 & 40$^{\circ}$ & 0.53 & 4.1 \\
1000& 40$^{\circ}$ & 0.20 & 1.5 \\
\hline
\end{tabular}
\caption{Simultaneous 90\% CL limits on dipole couplings obtainable from
$A_{ch}$ and
$A_{fb}$ for $\sqrt{s}=500$ GeV with integrated luminosity 200 fb$^{-1}$
and for $\sqrt{s}=1000$ GeV with integrated luminosity 1000 fb$^{-1}$ for
unpolarized beams. Cut-off $\theta_0$ is chosen to
optimize the sensitivity.}
\end{center}
\end{table}

Table 2 shows possible 90\% CL limits for the unpolarized case, when results
from $A_{ch}$ and $A_{fb}$ are combined.
The idea is that each asymmetry measures a different linear
combination of Im$c_d^{\g}$ and
Im$c_d^Z$. So a null result for the two asymmetries will correspond to two
different bands of regions allowed at 90\% CL in the space of Im$c_d^{\g}$ and
Im$c_d^Z$. The overlapping region of the two bands leads to the limits given in
Table 2. In this case, for 90\% CL, the asymmetry is required to be
$2.15/\sqrt{N}$, corresponding to two degrees of freedom. Incidentally, the same
procedure followed for $P_e=\pm 0.9$ gives much worse limits. 

\begin{table}
\begin{center}
\begin{tabular}{|c|c|c|c|c|c|c|}
\hline
&
\multicolumn{3}{|c|}{$A_{ch}$}&\multicolumn{3}{|c|}{$A_{fb}$}\\
\cline{2-7}
$\sqrt{s}$ (GeV)&  $\theta_0$ & Im$c_d^{\gamma}$ &
 Im$c_d^Z$ &$\theta_0$ & Im$c_d^{\gamma}$ & Im$c_d^Z$ \\
\hline
500 & 64$^{\circ}$ & 0.11 & 0.20 & $10^{\circ}$ & 0.11 & 0.20 \\
1000& 64$^{\circ}$ & 0.037& 0.069 &$10^{\circ}$ & 0.040 & 0.076 \\
\hline
\end{tabular}
\caption{Simultaneous limits on dipole couplings combining data from
polarizations $P_e=0.9$ and $P_e=-0.9$, using separately $A_{ch}$ and
$A_{fb}$ for $\sqrt{s}=500$ GeV with integrated luminosity 200 fb$^{-1}$
and for $\sqrt{s}=1000$ GeV with integrated luminosity 1000 fb$^{-1}$.
Cut-off $\theta_0$ is chosen to optimize the sensitivity.}
\end{center}
\end{table}

Similarly, using one of the two asymmetries, but two different polarizations of
the electron beam, one can get two bands in the parameter plane, which give
simultaneous limits on the dipole couplings. The results for electron
polarizations $P_e=\pm 0.9$ are given in Table 3 for each of the asymmetries
$A_{ch}$ and $A_{fb}$.

\begin{table}
\begin{center}
\begin{tabular}{|c|c|c|c|c|c|c|c|}
\hline
&&
\multicolumn{3}{|c|}{$A_{ud}$}&\multicolumn{3}{|c|}{$A_{lr}$}\\
\cline{3-8}
$\sqrt{s}$ (GeV) &$P_e$ & $\theta_0$ & Re$c_d^{\gamma}$ &
 Re$c_d^Z$ &$\theta_0$ & Im$c_d^{\gamma}$ & Im$c_d^Z$ \\
\hline
    &  0    &25$^{\circ}$ & $ 0.10$ & $ 0.034$ & $30^{\circ}$ & 0.024 & 0.14\\
500 &$+0.9$ &30$^{\circ}$ & $ 0.025$ & 0.037  & $35^{\circ}$ & 0.024 & 0.050\\
    &$-0.9$ &25$^{\circ}$ & $ 0.022$ & 0.032 & $30^{\circ}$ & 0.023 &0.038\\
\hline
    & 0      &30$^{\circ}$ & $ 0.029$ & $ 0.0096$ & $60^{\circ}$ &0.021&0.13 \\
1000& $+0.9$ &35$^{\circ}$ &  0.0068 & 0.010 &   $60^{\circ}$ &0.021&$ 0.045$\\
    & $-0.9$ &30$^{\circ}$ &  0.0061& 0.0089 &  $60^{\circ}$ &0.021&$ 0.035$\\
\hline
\end{tabular}
\caption{Individual 90\% CL limits on dipole couplings obtainable from
$A_{ud}$ and
$A_{lr}$ for $\sqrt{s}=500$ GeV with integrated luminosity 200 fb$^{-1}$
and for $\sqrt{s}=1000$ GeV with integrated luminosity 1000 fb$^{-1}$ for
different electron beam polarizations $P_e$. Cut-off $\theta_0$ is chosen to
optimize the sensitivity.}
\end{center}
\end{table}

Table 4 lists the 90\% CL limits which may be obtained on the real and
imaginary parts of the dipole couplings using $A_{ud}$ and $A_{lr}$, assuming
one of the couplings to be nonzero at a time.

\begin{table}\label{t5}
\begin{center}
\begin{tabular}{|c|c|c|c|c|c|c|}
\hline
&
\multicolumn{3}{|c|}{$A_{ud}$}&\multicolumn{3}{|c|}{$A_{lr}$}\\
\cline{2-7}
$\sqrt{s}$ (GeV)&  $\theta_0$ & Re$c_d^{\gamma}$ &
 Re$c_d^Z$ &$\theta_0$ & Im$c_d^{\gamma}$ & Im$c_d^Z$ \\
\hline
500 & 25$^{\circ}$ & 0.031& 0.045& $35^{\circ}$ & 0.031& 0.056\\
1000& 30$^{\circ}$ & 0.0085& 0.013 &$60^{\circ}$ & 0.028 & 0.052 \\
\hline
\end{tabular}
\caption{Simultaneous limits on dipole couplings combining data from
polarizations $P_e=0.9$ and $P_e=-0.9$, using separately $A_{ud}$ and
$A_{lr}$ for $\sqrt{s}=500$ GeV with integrated luminosity 200 fb$^{-1}$
and for $\sqrt{s}=1000$ GeV with integrated luminosity 1000 fb$^{-1}$. Cut-off
$\theta_0$ is chosen to optimize the sensitivity.}
\end{center}
\end{table}

Table 5 shows simultaneous limits on Re$c_d^{\g}$ and Re$c_d^Z$ obtainable from combining the data on $A_{ud}$ for $P_e=+0.9$ and $P_e=-0.9$, and similarly,
limits on Im$c_d^{\g}$ and Im$c_d^Z$ from data on $A_{lr}$ for the two
polarizations.

\section{Conclusions}

We have presented in analytic form  the single-lepton angular
distribution in the production and subsequent decay of $\tt$ in the
presence of electric and weak dipole form factors of the top quark. We have
included ${\cal O}(\alpha_s)$ QCD corrections in SGA. Anomalous contributions
to the $tbW$ decay vertex do not affect these distributions. 
We have also included effects of longitudinal electron beam polarization of 
90\%, while assuming the positron beam to be unpolarized. We have
then obtained analytic expressions for certain 
simple $CP$-violating polar-angle  
asymmetries, specially chosen so that they do not require the
reconstruction of the  $t$ or $\o t$ directions or energies. 
We have also evaluated numercially azimuthal asymmetries which need minimal
information on the $t$ or \tbar momentum direction alone.
We have
analyzed these asymmetries to obtain simultaneous 90\% CL limits on
the electric and weak dipole couplings which
would be possible at future linear $\ee$ collider like the proposed JLC 
operating at
$\sqrt{s}= 500$ GeV with an integrated luminosity of 200 fb$^{-1}$, and at 
$\sqrt{s}= 1000$ GeV with an integrated luminosity of 1000 fb$^{-1}$.

In general, simultaneous 90\% CL limts on $\cdg$ and $\cdz$ 
which can be obtained
with the polarized 500 GeV option are of the order of 0.1--0.2,
corresponding to dipole moments of about $(1-2)\times 10^{-17} e$ cm, 
if the asymmetries $A_{ch}$
or $A_{fb}$ are used. The limits improve by a factor of 3 or 4 if the azimuthal
asymetries $A_{ud}$ or $A_{lr}$ are used. However, putting in a top detection
efficiency factor of 10\% in the case of azimuthal asymmetries, where top
direction needs to be determined, would bring down these limts to the same
level of $(1-2)\times 10^{-17} e$ cm.

For $\sqrt{s}=1000$ GeV and an integrated luminosity of 1000 fb$^{-1}$, the
limits obtainable would be better by a factor of 3 or 4 in each case, bringing
them to the level of $(2-3)\times 10^{-18} e$ cm in the best cases.

Our general conclusion is that the sensitivity to the measurement of
individual dipole couplings Re$\cdg$ and Im$\cdz$ 
is improved considerably if the electron beam is
polarized, a situation which might easily obtain at linear colliders.
As a consequence, simultaneous  limits on all the couplings are improved by
beam polarization. 

The theoretical predictions for $c_d^{\g,Z}$ are at the level of
$10^{-2}-10^{-3} $, as for example, in the neutral-Higgs-exchange and
supersymmetric models of CP violation \cite{bern,asymm,new,sonirev}. 
In other  models, like the charged-Higgs-exchange \cite{sonirev} 
or third-generation 
leptoquark models \cite{pou4}, the prediction are even lower.  Hence
the measurements suggested here at the 500 GeV option 
cannot exclude these modes at the 90\% C.L.  It will be necessary to use the
1000 GeV option with a suitable luminosity to test at least some of the models.

It is necessary to repeat this study including experimental detection 
efficiencies.
Given an overall efficiency, we could still get an
idea of the limits on the dipole couplings by scaling them as the inverse square
root of the efficiency.

We have not included a cut-off on decay-lepton energies which may be required
from a practical point of view. However, our results are perfectly valid if the
cut-off is reasonably small. For example, for $\sqrt{s}=500$ GeV, the minimum
lepton energy allowed kinematically is about 7.5 GeV. So a cut-off below that
would need no modification of the results.

We have restricted ourselves to energies in the $t\o{t}$ continuum. Studies in
the threshold region are also interesting and have been investigated in
\cite{sumino}.

%\newpage
\vskip 1cm

\noindent{\Large \bf Appendix}
\vskip .5in

The expressions for $A_i$, $B_i$, $C_i$ and $D_i$ occurring in
equation (8) are listed below. They include  to first order the form factors
$\cdg$ and $\cdz$, as well as $c_M^{\g}$ and $c_M^Z$. Terms containing products
of $\cdgz$ with $c_M^{\g,Z}$ have been dropped. It is also understood that
terms proportional to products of $A$ or $B$ (which are of order $\alpha_s$)
and $\cdg$ or $\cdz$ have to omitted in the calculations.

\begin{eqnarray}
A_0 & = & 2 \left\{  (2-\beta^2) \left[ 2\vert c_v^{\g}\vert ^2 +
2(r_L+r_R){\rm Re}(c_v^{\g}c_v^{Z*}) + (r_L^2+r_R^2)\vert c_v^Z\vert ^2 \right]
\right.  \nonumber \\
&& \left.
+  \beta^2 (r_L^2+r_R^2)\vert c_a^Z\vert ^2
-2\beta^2\left[2{\rm Re}(c_v^{\g}c_M^{\g *})
\right. \right. \nonumber \\
&& \left. \left.
+(r_L+r_R)\Re (c_v^{\g}c_M^{Z*}+c_v^Zc_M^{\g *})
+ (r_L^2+r_R^2)\Re (c_v^Zc_M^{Z*})\right]
\right. (1- P_e P_{\overline e}) \nonumber \\
&& +  \left. (2-\beta^2) \left[
2(r_L-r_R)\Re (c_v^{\g}c_v^{Z*})
+ (r_L^2-r_R^2)\vert c_v^Z\vert ^2 \right]
\right.
\nonumber \\
& &  \left.
  +  \beta^2 (r_L^2-r_R^2)\vert c_a^Z\vert ^2
- 2\beta^2
\left[(r_L-r_R)\Re (c_v^{\g}c_M^{Z*}+c_v^Zc_M^{\g *})
\right. \right.  \nonumber \\
&& \left. \left.
 + (r_L^2 - r_R^2)\Re (c_v^Zc_M^{Z *})
\right] \right\} (P_{\overline e}-P_e), \nonumber \\
A_1&=& -8 \beta \Re \left( c_a^{Z*} \left\{
\left[(r_L-r_R)c_v^{\g} + (r_L^2-r_R^2)c_v^Z \right]  (1-
P_eP_{\overline e}) \right. \right. \nonumber \\ && \left. \left. +
\left[(r_L+r_R)c_v^{\g} + (r_L^2+r_R^2)c_v^Z \right] (P_{\overline
e}-P_e) \right\} \right),  \nonumber\\
A_2&=& 2 \beta^2 \left\{ \left[
2\vert c_v^{\g}\vert ^2 + 4\Re (c_v^{\g}c_M^{\g *})
+ 2(r_L+r_R)\Re (c_v^{\g}c_v^{Z*} +c_v^{\g}c_M^{Z*} + c_v^Zc_M^{\g *})
\right.\right. \nonumber \\
&&\left. \left.
+(r_L^2+r_R^2)\left(
\vert c_v^Z\vert ^2 + \vert c_a^Z\vert ^2 +2 \Re (c_v^Zc_M^{Z*}) \right)
 \right]  (1- P_e P_{\overline e})
\right. \nonumber \\ && \left. +  \left[ 2(r_L-r_R)\Re (c_v^{\g}c_v^{Z*}
+c_v^{\g}c_M^{Z*}
+ c_v^Z c_M^{\g*}) \right. \right. \nonumber \\
&& \left.\left. 
+(r_L^2-r_R^2)\left( \vert c_v^Z\vert ^2 + \vert c_a^Z\vert ^2
+ 2\Re (c_v^Zc_M^{Z*}) \right) \right]
(P_{\overline e}-P_e) \right\}, \nonumber\\
B_0^{\pm}&=& 4\beta
\left\{ \left(\Re c_v^{\g} + r_L\Re c_v^Z\right) \left(r_L \Re c_a^Z \mp {\rm
Im}c_d^{\g} \mp r_L {\rm Im}c_d^Z \right) (1-P_e)(1+P_{\overline e})
\right. \nonumber \\ 
&& \left. +  \left(\Re c_v^{\g} + r_R\Re c_v^Z\right)
\left(r_R \Re c_a^Z \mp {\rm Im}c_d^{\g} \mp r_R {\rm Im}c_d^Z \right)
(1+P_e)(1-P_{\overline e}) \right\}, \nonumber \\
B_1&=& -4 \left\{
\left[\vert c_v^{\g}+r_Lc_v^Z\vert ^2+ \beta^2 r_L^2 \vert c_a^Z\vert ^2 \right]
(1-P_e)(1+P_{\overline e}) \right. \nonumber \\ && \left.  - 
\left[\vert c_v^{\g}+r_Rc_v^Z\vert ^2+ \beta^2 r_R^2 \vert c_a^Z\vert ^2 \right]
(1+P_e)(1-P_{\overline e}) \right\}, \nonumber \\
B_2^{\pm}&=& 4\beta
\left\{  \left(\Re c_v^{\g} + r_L\Re c_v^Z\right) \left(r_L \Re c_a^Z \pm {\rm
Im}c_d^{\g} \pm r_L {\rm Im}c_d^Z \right) (1-P_e)(1+P_{\overline e})
\right. \nonumber \\ && \left.+ \left(\Re c_v^{\g} + r_R\Re c_v^Z\right)
\left(r_R\Re  c_a^Z \pm {\rm Im}c_d^{\g} \pm r_R {\rm Im}c_d^Z \right)
(1+P_e)(1-P_{\overline e}) \right\}, \nonumber\\
C_0^{\pm}&=&4
\left\{ \left[ \vert c_v^{\g} + r_Lc_v^Z\vert ^2
- \beta^2 \gamma^2 \left( \Re c_v^{\g} + r_L \Re c_v^Z \right)\left( \Re
c_M^{\g} + r_L \Re c_M^Z \right) \right. \right. \nonumber \\
&&\left. \left.   \pm \beta^2
\gamma^2 r_L \Re c_a^Z \left( {\rm Im} c_d^{\g} + {\rm Im} c_d^Z
r_L \right) \right] (1-P_e)(1+P_{\overline e}) \right. \nonumber \\
&& \left. - \left[ \vert c_v^{\g} + r_Rc_v^Z\vert ^2
- \beta^2 \gamma^2 \left( \Re c_v^{\g} + r_R \Re c_v^Z \right)\left( \Re
c_M^{\g} + r_R \Re c_M^Z \right) \right. \right. \nonumber \\
&& \left. \left. \pm \beta^2
\gamma^2 r_R \Re c_a^Z \left( {\rm Im} c_d^{\g} + {\rm Im} c_d^Z
r_R \right) \right] (1+P_e)(1-P_{\overline e}) \right\}, \nonumber\\
C_1^{\pm}&=&- 4\beta \left\{ \left[ \left(\Re c_v^{\g} +
r_L\Re c_v^Z\right) \left(r_L \Re c_a^Z \pm \gamma^2 {\rm Im}c_d^{\g} \pm r_L
\gamma^2 {\rm Im}c_d^Z \right) \right. \right. \nonumber \\ &&\left. \left.
- \beta^2 \gamma^2 r_L \Re c_a^Z\left(\Re c_M^{\g} +r_L \Re c_M^Z) \right)
\right] (1-P_e)(1+P_{\overline e}) \right.
\nonumber \\
&& \left.+ \left[ \left(\Re c_v^{\g} + r_R\Re c_v^Z\right)
\left(r_R \Re c_a^Z \pm \gamma^2 {\rm Im}c_d^{\g} \pm r_R
\gamma^2 {\rm Im}c_d^Z \right) \right. \right. \nonumber \\ && \left. \left.
- \beta^2 \gamma^2 r_R \Re c_a^Z\left(\Re c_M^{\g} +r_R \Re c_M^Z) \right)
\right] (1+P_e)(1-P_{\overline e})
\right\},\nonumber \\
D_0^{\pm}&=& 4 \beta \left\{ \left[
\Im \left[\left( (c_V^{\g} +
r_L c_v^Z) -\beta^2 \gamma^2 ( c_M^{\g} + r_L c_M^Z)\right) r_L c_a^{Z*}\right]  \right. \right.
\nonumber \\ && \left. \left.
\mp  \gamma^2 \left(\Re c_v^{\g} +
r_L\Re c_v^Z\right) \left( {\rm Re}c_d^{\g} + r_L {\rm Re}c_d^Z \right)
\right]
(1-P_e)(1+P_{\overline e}) \right. \nonumber \\ && \left.
-\left[ \Im \left[\left( (c_V^{\g} +
r_R c_v^Z) -\beta^2 \gamma^2 ( c_M^{\g} + r_R c_M^Z)\right) r_R c_a^{Z*}\right]  \right.\right.
\nonumber \\ &&\left. \left.
\mp \gamma^2 \left(\Re c_v^{\g} + r_R\Re c_v^Z\right) \left({\rm Re}c_d^{\g} 
+ r_R {\rm Re}c_d^Z \right)\right]
 (1+P_e)(1-P_{\overline e}) \right\}, \nonumber \\
D_1^{\pm}&=& 4 \beta^2  \gamma^2 \left\{ \left[ (\Re c_v^{\g} + r_L \Re c_v^Z)(\Im
c_M^{\g}  \right. \right. \nonumber \\
&&\left. \left. + r_L \Im c_M^Z) \pm r_L \Re c_a^Z \left( {\rm Re}c_d^{\g} +
r_L {\rm Re} c_d^Z \right)\right] (1-P_e)(1+P_{\overline e})
\right. \nonumber \\
&& \left.+\left[ (\Re c_v^{\g} + r_R \Re c_v^Z)(\Im c_M^{\g} + r_R \Im c_M^Z) 
\right.\right.
\nonumber \\
&& \left.\left. \pm r_R \Re c_a^Z \left( {\rm Re}c_d^{\g} 
+ r_R {\rm Re} c_d^Z \right)\right]
(1+P_e)(1-P_{\overline e})
\right\}. \nonumber
\end{eqnarray}
Use has been made of
\[
r_L =  \f{(\half - x_W)}{(1-m_Z^2/s)\sqrt{x_w (1-x_w)}} 
\]
and 
\[
r_R =  \f{- x_W}{(1-m_Z^2/s)\sqrt{x_w (1-x_w)}} 
\]
in writing the above equation.

\end{document}